# Non-ideal gas dynamics under confinement: rarefaction effect, dense effect and molecular interaction

Baochao Shan, Peng Wang, Songze Chen, Zhaoli Guo [*]

*State Key Laboratory of Coal Combustion, Huazhong University of Science and Technology, Wuhan, Hubei 430000, China*

\* Corresponding e-mail address: *zlguo@hust.edu.cn*

**Abstract:** The effects of volume exclusion and long-range intermolecular attraction are investigated by the simplified kinetic model for surface-confined inhomogeneous fluids. Gas dynamics of the ideal gas, the hard-sphere fluid and the real gas are simulated by the Boltzmann equation, the Enskog equation and the simple kinetic equation, respectively. Only the Knudsen minimum appears for the ideal gas, while both the Knudsen minimum and the Knudsen maximum occur for the hard-sphere fluid and the real gas under certain confinements, beyond which the maximum and minimum may disappear. The Boltzmann equation and the Enskog equation overestimates and underestimates the mass flow rate of the real gas dynamics under confinement, respectively, where the volume exclusion and the long-range intermolecular attractive potential among molecules are not ignorable. With the increase of the channel width, gas dynamics of the hard-sphere fluid and the real gas tends to the Boltzmann prediction gradually. The density inhomogeneity, which hinders the flow under confinement, is more obvious when the solid fraction is larger. The anomalous slip occurs for real gas under constant confinement. The flow at a smaller Knudsen number (larger solid fraction or channel width) contributes more practical amount of mass transfer, although the rarefaction effects is more prominent at larger Knudsen numbers. The temperature has no effect on density and velocity profiles of the ideal gas and the hard-sphere fluid, but the energy parameter among the real gas molecules decreases with the increasing temperature and the real gas dynamics tends to the hard-sphere ones consequently.

**Introduction**

Non-ideal gas flow in micropores or nanopores is encountered in many engineering applications, such as natural gas production and geological storage of carbon dioxide [1-3]. In these applications, gas behaviors deviate from ideal gas ones significantly due to the high pressure condition, namely the non-ideal gas effect [4]. Understanding the non-ideal gas dynamics under tight confinements is central to optimize the design of these engineering problems.

The equation of state (EOS) of real gases, such as the van der Waals, Soave-Redlich-Kuang and Peng-Robinson, is commonly used to account for the non-ideal gas effect when studying gas flow behaviors in natural gas reservoirs [5-7]. Comparing to the ideal gas EOS, gas molecule size and molecular interaction are taken into account in the real gas EOS. Another way to consider the non-ideal gas effect is to introduce gas compressibility factor [8, 9], which measures the deviation degree of real gas behaviors from ideal gas ones. For example, the gas compressibility factor is usually employed to calculate the gas mean free path and mass flow rate in shale gas reservoirs [10, 11]. However, it fails to describe the change of gas properties, such as the density and viscosity, across the flow path.

The non-ideal gas flow behaviors and rarefaction effects have been widely studied by various approaches, such as the pore-scale simulation [12, 13], pore network modeling [6, 14], lattice Boltzmann method (LBM) [15-17] and empirical modification on the Darcy's law [10, 18-20]. However, the simultaneous consideration of non-ideal gas effect and rarefaction effect is still not effective. For example, the standard LBM fails to properly describe the rarefaction effects, although it can be extended to the slip flow regime [13]; accounting for the non-ideal gas effect and rarefaction effect by the EOS of real gases and a velocity slip at the boundary separately is empirical and fails to obtain the flow details. The coupling of the high $Kn$ and non-ideal effects was investigated by adopting the Klinkenberg modification to account for the rarefaction effect and introducing the characteristic pressure and viscosity of real gases to account for the non-ideal effect [1], but it is still phenomenological and solving the model by the LBM limits its application in more complicated flows. Meanwhile, the rarefaction effect not only exhibits as slippage at fluid-solid interface, but also changes the velocity distribution across the flow path [21]. Therefore, a more consistent procedure is needed to consider non-ideal gas effect and rarefaction effect for gas flow under confinement simultaneously.

From a microscopic perspective, the non-ideal gas effects arise from the non-ignorable molecule size and molecular interaction [4]. As we know, the Boltzmann equation is applicable to all flow regimes, from the continuum to the free molecular flow regimes [22], given that the gas is dilute [23]. Therefore, it can capture the rarefaction effects of the whole system without empirical modification at the boundary. However, the Boltzmann equation recovers to the ideal gas EOS and only describes ideal gas dynamics. The Enskog equation takes the gas molecule size into account and describes flow behavior of hard-sphere fluids [24], where the long-range intermolecular attraction is not considered. Adopting the non-ideal gas LBM model, Shi et al. [25] studied the dense gas flow behaviors in micro-channels using a second-order slip boundary condition. Based on the non-ideal gas model, shale gas flow mechanisms were characterized and analyzed by Ren et al. [15] during gas production process using the bounce-back and specular-reflection boundary condition. Recently, the non-equilibrium gas dynamics of dense gases under confinement were investigated by the generalized Enskog equation for two dimensional hard disks [26]. However, the long-range intermolecular attraction were not considered, the importance of which has been demonstrated in the van der Waals theory for liquids [27, 28]. The Enskog-Vlasov equation considers both the gas molecule size and molecular interaction, and recovers to the real gas EOS of van der Waals. Consequently, it accounts for the rarefaction and non-ideal gas effects simultaneously in a self-consistent way [28]. The Enskog-Vlasov equation has been successfully applied to study gas-liquid phase transitions [29] and net condensation/evaporation [30, 31]. However, the Enskog-Vlasov equation is for a homogeneous system. To account for the inhomogeneity, the local average density method (LADM) [32] needs to be introduced, as in the previous kinetic model [33], where the wall potential, cross-sectional fluid inhomogeneity, volume exclusion and intermolecular attraction were simultaneously considered. The aim of this paper is to investigate the effects of volume exclusion and intermolecular attraction in gas dynamics in a surface-confined inhomogeneous system and thus the wall potential will not be considered for simplification.

In this paper, the gas dynamics of ideal gas, hard-sphere fluids and real gas are simulated by the Boltzmann equation, the Enskog equation and the simple kinetic equation [33], respectively. The purpose of this paper is to compare the dynamics of three different fluids and investigate the effects of volume exclusion and attractive forces among molecules on flow behaviors.

## 1 The simple kinetic equation of real gases

### 1.1 Kinetic equation

In the kinetic model of Davis [34], the finite size of fluid molecules and the volume exclusion effect are considered by the hard-sphere potential, and the long-range intermolecular attraction is coupled by the mean-field theory [35], which can be written as [36]

$$\partial_t f + \boldsymbol{\xi} \cdot \boldsymbol{\nabla}_r f - m^{-1}\boldsymbol{\nabla}_r \phi_{ext} \cdot \boldsymbol{\nabla}_\xi f = \Omega_E + m^{-1}\boldsymbol{\nabla}_\xi f \cdot \int n(\boldsymbol{r}_1)\chi(\boldsymbol{r},\boldsymbol{r}_1)\boldsymbol{\nabla}_r \phi_{att} d\boldsymbol{r}_1, \qquad (1)$$

where $f(\boldsymbol{r}, \boldsymbol{\xi}, t)$ is the velocity distribution function of molecular velocity $\boldsymbol{\xi}$ at spatial position $\boldsymbol{r}$ and time $t$, $\boldsymbol{r}_1$ is also the spatial position, $\partial_t$ represents partial derivative in terms of time $t$, $m$ is the molecular mass, $n$ is the number density, $\chi$ is the radial distribution function, $\boldsymbol{\nabla}_r$ and $\boldsymbol{\nabla}_\xi$ represent gradient operators in terms of space $\boldsymbol{r}$ and velocity $\boldsymbol{\xi}$, respectively; $\phi_{ext}$ and $\phi_{att}$ are the external and intermolecular attractive potential, respectively; and $\Omega_E$ is the Enskog collision operator for hard-sphere fluids.

The Enskog collision operator $\Omega_E$ is such complicated that it is difficult in practical applications, which needs further simplification by projecting it into the Boltzmann collision part $\Omega_B$ and the excess part $\Omega_{ex}$ [33], which has been proved as an effective way to treat the original complicated collision operator [21, 33]. For an isothermal case, the Boltzmann collision operator $\Omega_B$ can be approximated by the Bhatnagar-Gross-Krook (BGK) operator. In order to apply the kinetic equation (1) to inhomogeneous system, Guo et al. [33] proposed an expression for the excess collision part $\Omega_{ex}$ accounting for the volume exclusion effect, which can be approximated as [33]

$$\Omega_{ex} = -V_0 f^{eq} (\boldsymbol{\xi} - \boldsymbol{u}) \cdot \left[2\boldsymbol{A}\chi(\overline{n}) + \boldsymbol{B}\overline{n}\right], \qquad (2)$$

where $V_0 = 2\pi\sigma^3/3$, $\sigma$ is the effective molecular diameter, $\boldsymbol{u}$ is the flow velocity, $\overline{n} = \int w(\boldsymbol{r}_1) n(\boldsymbol{r} + \boldsymbol{r}_1) d\boldsymbol{r}_1$ is the average density of $n$ with a weight function $w(\boldsymbol{r}_1)$ [37, 38], and $\boldsymbol{A}$ and $\boldsymbol{B}$ can be expressed as [33, 39]

$$\boldsymbol{A} = \frac{120}{\pi\sigma^5} \int_{|\boldsymbol{r}_1-\boldsymbol{r}|<\sigma/2} (\boldsymbol{r}_1 - \boldsymbol{r})\overline{n}(\boldsymbol{r}_1) d\boldsymbol{r}_1, \qquad (3)$$

and

$$B = \frac{120}{\pi\sigma^5} \int_{|r_1-r|<\sigma/2} (r_1 - r) \chi(r_1) dr_1. \tag{4}$$

Meanwhile, the effective molecular diameter can be approximated by [40]

$$\sigma \approx \frac{1 + a_1 T_r}{1 + a_2 T_r + a_3 T_r^2}, \tag{5}$$

where $a_1 = 0.2977$, $a_2 = 0.33163$, $a_3 = 0.00104771$; and $T_r = k_B T/\varepsilon$ is the reduced temperature, with $k_B$ being the Boltzmann constant, $T$ being the system temperature and $\varepsilon$ being the energy parameter of the fluids.

For the long-range attractive potential, i.e., the last term in Eq.(1), the radial distribution function is approximately unity in the attractive range [33] and the term can be rearranged as

$$J = m^{-1} \nabla_\xi f \cdot \nabla_r \left[ \int \bar{n}(r_1) \phi_{att}(r_1 - r) dr_1 \right], \tag{6}$$

where $\phi_{att}$ is the attractive part of the molecular interaction, which acts only in the attractive range, i.e., $|r_1 - r| > \sigma$. Generally, it can be represented by the 12 – 6 LJ potential among molecules as

$$\phi_{att} = 4\varepsilon \left[ \left(\frac{\sigma}{r}\right)^{12} - \left(\frac{\sigma}{r}\right)^6 \right], \quad r > \sigma. \tag{7}$$

Taking all the above analysis into account, the kinetic equation (1) is transformed into the following simple kinetic equation proposed by Guo et al.[33]

$$\partial_t f + \xi \cdot \nabla_r f - m^{-1} \nabla_r \phi_{ext} \cdot \nabla_\xi f = -\tau^{-1}(f - f^{eq}) + \Omega_{ex} + J, \tag{8}$$

where the volume exclusion effect is accounted for by $\Omega_{ex}$ and the long-range molecular interaction is accounted for by the term $J$, which can be calculated by Eqs.(2) and (6), respectively.

**1.2 Analysis and comparison of kinetic equations**

(1) The simple kinetic equation (8) considers the effects of intermolecular attractive forces by the mean-field theory, which is an extension of the Enskog-Vlasov equation from homogeneous to inhomogeneous system. If the parameter $\eta$ calculated from the local density is smaller than the packing limit and the system is weak inhomogeneous, then we have [39]

$$\bar{n} \approx n, \boldsymbol{A} \approx \boldsymbol{\nabla} n, \boldsymbol{B} \approx \boldsymbol{\nabla} \chi, \tag{9}$$

and the model becomes that proposed in He et al. [41].

(2) The Enskog-BGK equation: if the long-range attractive tail is not considered, the simple kinetic equation (8) is transformed into the Engkog-BGK equation as

$$\partial_t f + \boldsymbol{\xi} \cdot \boldsymbol{\nabla}_r f - m^{-1} \boldsymbol{\nabla}_r \phi_{ext} \cdot \boldsymbol{\nabla}_{\boldsymbol{\xi}} f = -\tau^{-1}\left(f - f^{eq}\right) + \Omega_{ex}, \tag{10}$$

where the Enskog collision operator is projected into a Boltzmann collision term and an excess collision term, as introduced above, and the finite size and the non-local collisions among molecules are explicitly included by the excess collision term $\Omega_{ex}$. Later, we will show that the Enskog-BGK model produces results in quantitatively agreement with the full Enskog equation ones.

(3) The Boltzmann-BGK equation: if the gas is dilute and the molecular size can be ignored, the excess collision term can be removed from Eq.(10), and the kinetic equation is transformed into the Boltzmann-BGK equation as

$$\partial_t f + \boldsymbol{\xi} \cdot \boldsymbol{\nabla}_r f - m^{-1} \boldsymbol{\nabla}_r \phi_{ext} \cdot \boldsymbol{\nabla}_{\boldsymbol{\xi}} f = -\tau^{-1}\left(f - f^{eq}\right), \tag{11}$$

where $\chi = Y = 1$.

As introduced above, the Boltzmann equation only retains the thermodynamic properties of a perfect gas [42]. Therefore, in the Boltzmann-BGK equation, any inclusion of dense effect only changes the value of the numerical viscosity, if the finite size of molecules and the volume exclusion effect are not considered explicitly [43].

Through the Chapman-Enskog analysis, the following Navier-Stokes hydrodynamic equations can be recovered from Eqs.(8), (10) or (11)

$$\partial_t \rho + \boldsymbol{\nabla} \cdot \left(\rho \boldsymbol{u}\right) = 0, \tag{12}$$

$$\partial_t \left(\rho \boldsymbol{u}\right) + \boldsymbol{\nabla} \cdot \left(\rho \boldsymbol{u}\boldsymbol{u}\right) = -\boldsymbol{\nabla} p + \boldsymbol{\nabla} \cdot \left\{\mu \left[\boldsymbol{\nabla} \boldsymbol{u} + \left(\boldsymbol{\nabla} \boldsymbol{u}\right)^T\right]\right\} - n \boldsymbol{\nabla}_r \phi_{ext}. \tag{13}$$

Note that when obtaining the Eqs.(12) and (13) from the simple kinetic equation (8), the approximation in Eq.(9) is adopted and the capillary tensor term is ignored, since no interface appears in our studied system.

Although the same hydrodynamic equations can be obtained from the Eqs.(8), (10) and (11), the hydrodynamic pressure in Eq.(13) is different. For the Boltzmann-BGK equation of Eq.(11), the EOS satisfies

$$p = nk_B T, \tag{14}$$

which is the EOS of an ideal gas. Therefore, the fluid in the Boltzmann-BGK equation is the ideal gas.

For the Enskog-BGK equation, the EOS satisfies

$$p = nk_B T \left(1 + n V_0 \chi\right), \tag{15}$$

which considers the finite size of fluid molecules, but ignores the effects of fluid intermolecular attraction. Therefore, the fluid in the Enskog-BGK equation is the hard-sphere fluid.

For the simple kinetic equation (8), the EOS satisfies

$$p = nk_B T \left(1 + n V_0 \chi\right) - an^2. \tag{16}$$

where the parameter $a$ accounts for the attractive potential among molecules. In Eq.(16) both the molecule size and molecular interaction are considered. Therefore, the fluid in the simple kinetic equation is the non-ideal gas or real gas.

For an inhomogeneous system, the relaxation time $\tau$ is calculated as [24]

$$\tau = \frac{5.0}{16n\sigma^2}\sqrt{\frac{1}{\pi k_B T}}\bar{n}V_0\left(Y^{-1} + 0.8 + 0.7614Y\right), \tag{17}$$

where

$$Y = \bar{n}V_0\chi(\bar{n}), \quad \chi(\bar{n}) = \frac{1-0.5\eta}{(1-\eta)^3}, \quad \eta = \frac{\bar{n}V_0}{4}, \quad V_0 = \frac{2\pi\sigma^3}{3}. \tag{18}$$

where $\chi$ is the radial distribution function, and $\eta$ is the solid fraction, reflecting the denseness of the fluids. The relaxation time in Eq.(17) recovers to the normal expression for a homogeneous system.

In the rest of the paper, we will clarify the role of repulsive and attractive forces among molecules and compare gas dynamics of ideal gas, hard-sphere fluids and real gas by comparing the Boltzmann-BGK equation, the Enskog-BGK equation and the simple kinetic equation.

**2 Model solution**

In this paper, the above models are solved by the DUGKS, which has the characteristics of asymptotic preserving, low dissipation, second order accuracy and multidimensional nature, under the constraint of $\int_0^W n dy = n_0$, where $n_0$ is the pore averaged density and $W$ is the effective channel width ($L$ for the ideal gas or $H$ for the hard-sphere fluids and the real gas, as shown in Figure 1). The number density $n$ and the flow velocity $\boldsymbol{u}$ can be calculated by taking the moments of the distribution function

$$n = \int f d\boldsymbol{\xi}, \tag{19}$$

and

$$\boldsymbol{u} = n^{-1} \int \boldsymbol{\xi} f d\boldsymbol{\xi}. \qquad (20)$$

The detailed procedures of solving the model can be referred to our previous studies [21, 44, 45].

In this paper, the diffuse boundary condition is adopted, which is

$$f\left(\boldsymbol{r}_w, \boldsymbol{\xi}_i\right) = f^{eq}\left(\boldsymbol{\xi}_i; \rho_w, \boldsymbol{u}_w\right), \ \boldsymbol{\xi}_i \cdot \boldsymbol{n} > 0, \qquad (21)$$

where the $\rho_w$ is the density at the wall determined by the condition that no particles can go through the wall [45], *i.e.*,

$$\rho_w = -\sum_{\boldsymbol{\xi} \cdot \boldsymbol{n} < 0} \left(\boldsymbol{\xi} \cdot \boldsymbol{n}\right) f\left(\boldsymbol{r}_w, \boldsymbol{\xi}_i\right) \Big/ \sum_{\boldsymbol{\xi} \cdot \boldsymbol{n} > 0} \left(\boldsymbol{\xi} \cdot \boldsymbol{n}\right) f^{eq}\left(\boldsymbol{\xi}_i; 1, \boldsymbol{u}_w\right), \qquad (22)$$

## 3 Numerical tests and discussion

The introduced three models above are compared and analyzed in this section to elucidate the gas dynamics confined in two paralleled plates under different conditions, where the system is chosen as methane at the temperature of 363 K. As shown in Figure 1, the effective channel width in the Boltzmann-BGK equation is $L$, while it is $H$ in the Engkog-BGK equation and the simple kinetic equation, since the molecular size is considered in the Enskog theory. The fluid is moving in the $x$ direction under a small external acceleration as $a = 0.001268$.

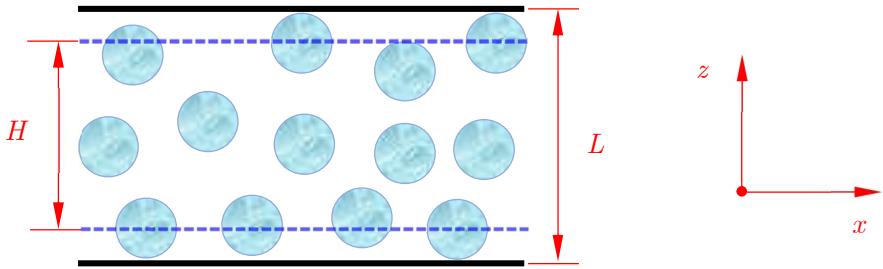

Figure 1: Effective channel width $L$ for Boltzmann-BGK model, and $H$ for the Enskog-BGK equation and the simple kinetic equation, where $L = H + \sigma$.

The non-dimensional number density $n$, velocity $u_n$ (or slip velocity $u_{sn}$), momentum $M_n$ and mass flow rate $Q_n$ are, respectively, defined as

$$n_n = \frac{n}{n_0}, \tag{23}$$

$$u_n = \frac{u}{aW/\sqrt{2k_BT/m}}, \quad u_{sn} = \frac{u_s}{aW/\sqrt{2k_BT/m}} \tag{24}$$

$$M_n = n_n u_n = \frac{nu}{n_0 aW/\sqrt{2k_BT/m}}, \tag{25}$$

$$Q_n = \frac{\int_0^W nu\, dy}{n_0 aW^2/\sqrt{2k_BT/m}} \tag{26}$$

As shown in Figure 2, the Knudsen minimum is captured for all three fluids, *i.e.*, the ideal gas, the hard-sphere fluid and the real gas at the $Kn \approx 0.86$. For the hard-sphere fluid and the real gas, it may disappear under tight confinement. For example, it disappears when $H/\sigma = 5$ for the hard-sphere fluid. Under tighter confinement, the effects of wall potential become greatly significant, leading to strong inhomogeneity in the system. Therefore, the simple diffuse boundary condition may fail to model the nature of the wall and the wall potential should be modeled separately. Consequently, the cases of tighter confinement ($H/\sigma < 5$) are not considered in this work. For $Kn > 1$, the $Q_n$ from the Boltzman equation (Ohwada) [46], the Boltzmann-BGK equation, the full Enskog model and the Enskog-BGK model almost converge to a single straight line, although it is slightly higher for the Boltzmann-BGK and the Enskog-BGK results. The difference is also found in previous studies [47]. Overall, the tightness barely affects the non-dimensional mass flow rate when $Kn > 1$. For $Kn < 1$, the results from the full Enskog equation and the Enskog-BGK equation also display a quantitatively good agreement (Figure 2a). A locally Knudsen maximum is observed when $Kn$ <0.1, which is $Kn \approx 0.019$ for $H/\sigma = 30$, $Kn \approx 0.029$ for $H/\sigma = 15$ and $Kn \approx 0.053$ for $H/\sigma = 10$. The same phenomenon is also observed in the two-dimensional Enskog equation [26].

The effects of the attractive forces among molecules on the mass flow rate can be observed in Figure 2 (b). All three models produce the same mass flow rate when $Kn > 1$, but it is severely underestimated by the hard-sphere fluids or overestimated by the ideal gas comparing to the real gas dynamics when $Kn < 1$. Considering the fact that the Knudsen number ranges from 0.0003 to 3 in typical shale gas reservoirs [5], and it is smaller in tight or conventional natural gas reservoirs, the real gas effect is necessary to be taken in account to study the gas dynamics under confinement, where the attractive forces play an important role in gas transport behaviors.

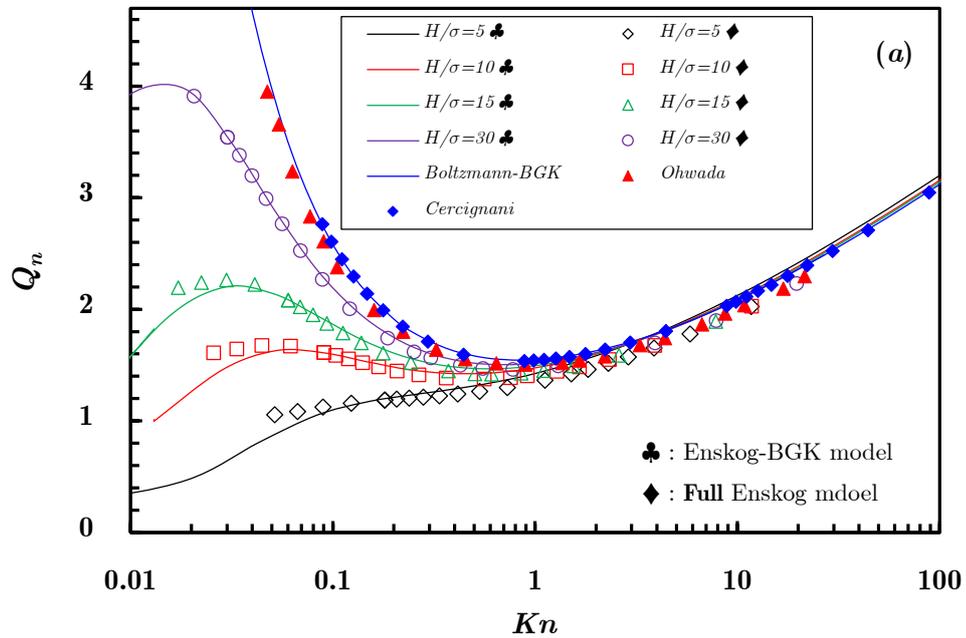

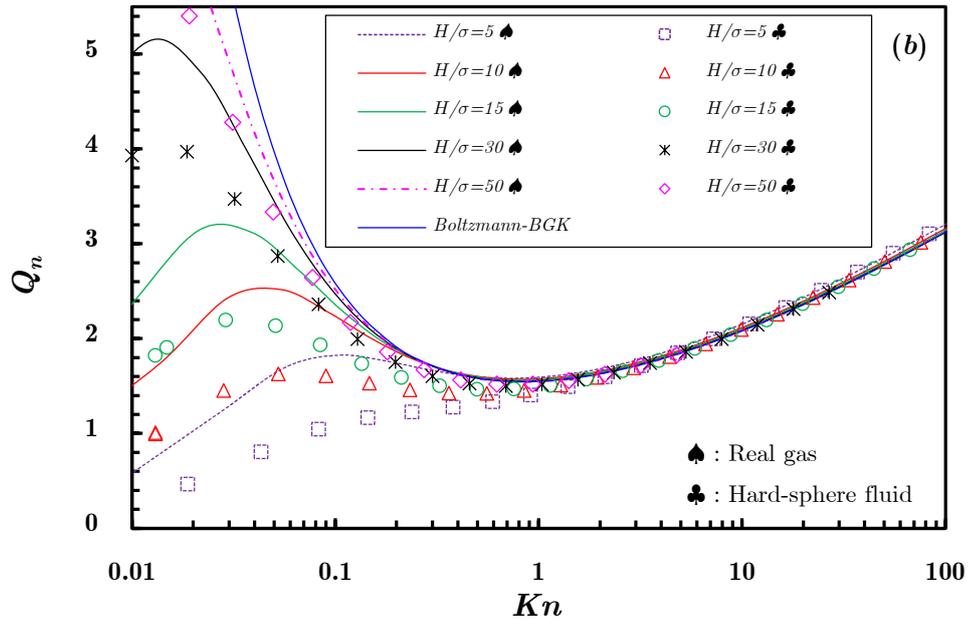

Figure 2: The non-dimensional mass flow rate of ideal gas, hard-sphere fluid and the real gas. The legend of Ohwada in (a) represents the results from the full Boltzmann equation [46], and the Cercignani represents the results from the Boltzmann-BGK equation [48].

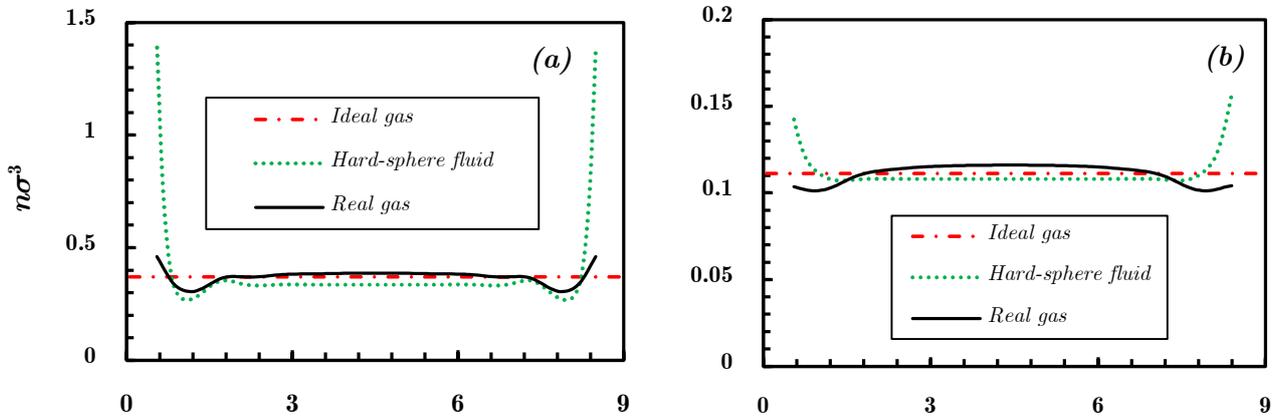

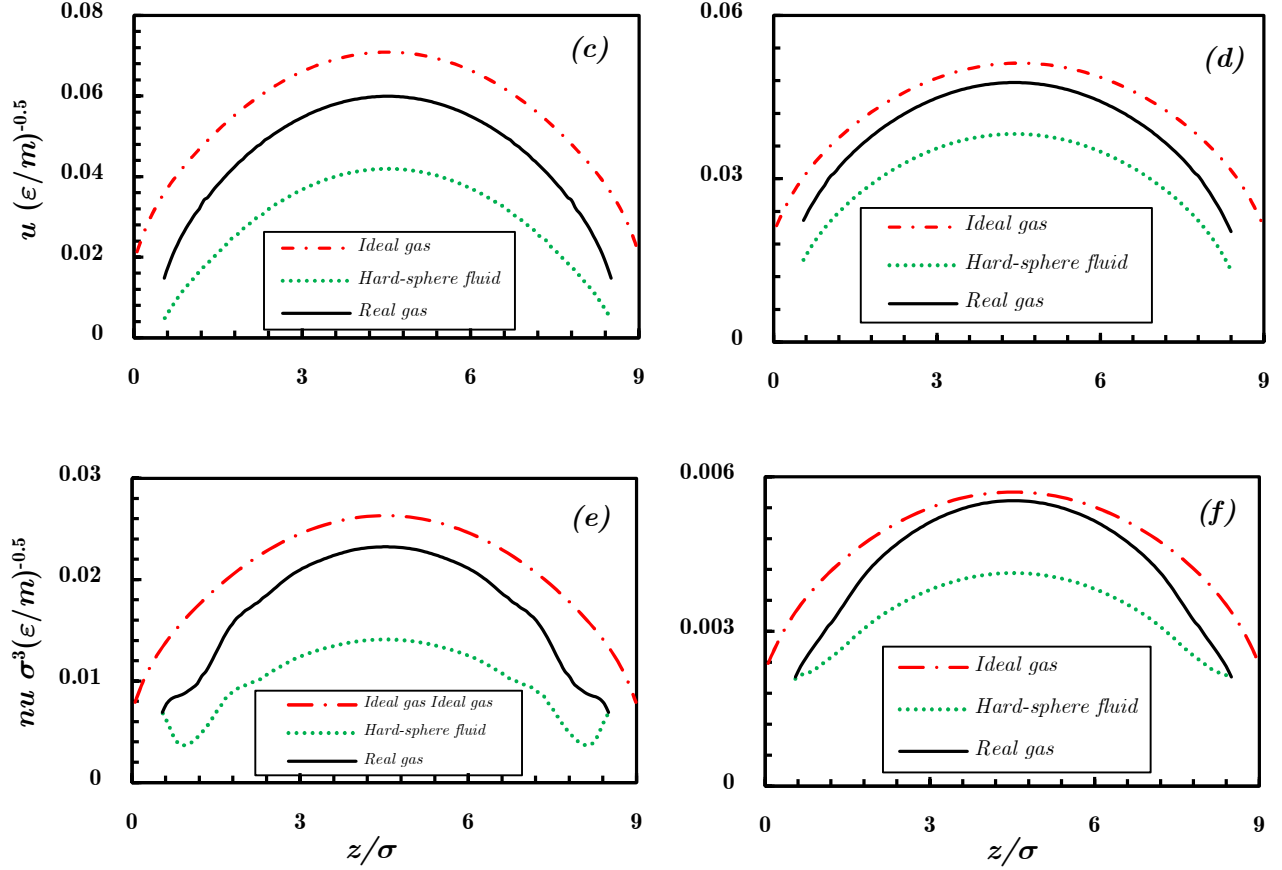

Figure 3: The density (*a* and *b*), velocity (*c* and *d*), and momentum (*e* and *f*) profiles of methane flow in slit channels of $H/\sigma = 8.04$ at the temperature of $T = 2.45\ \varepsilon/k_B$. The *a*, *c* and *e* are for profiles with the pore averaged density at $n_0 = 0.371\ \sigma^{-3}$ ($Kn = 0.0671$ for ideal gas, $Kn = 0.0389$ for hard-sphere fluid and real gas), and the *b*, *d* and *f* are for profiles with pore averaged density at $n_0 = 0.111\ \sigma^{-3}$ ($Kn = 0.2237$ for ideal gas, $Kn = 0.1924$ for hard-sphere fluid and real gas).

The density, velocity and momentum profiles across the confinement are shown in Figure 3. The density profile is unitary for ideal gases, and it is more oscillatory across the channel when the pore averaged density is larger for both the hard-sphere fluids and the real gases. At larger pore averaged density ($n_0 = 0.371\ \sigma^{-3}$), the molecules of hard-sphere fluids and real gas accumulate near the wall, and the density decreases with the distance from the wall increasing; a concave forms at approximately one or two diameters away from the wall on the density profile, after which the bulk region follows. At smaller pore averaged density ($n_0 =$ 0.111 $\sigma^{-3}$), no concave is observed for the hard-sphere fluids and the real gases. The real gas density near the wall is smaller than that of the bulk region, due to the action of the long-range intermolecular attraction. The intermolecular attractive forces works like an internal negative pressure, which pulls the molecules in a more stable state and makes the density distribution more unitary across the channel, as we can see from Figure 3 (*a*) and (*b*). Consequently, the mass flow rate or the flow velocity of the real gas under confinement is between that of the ideal gas and hard-sphere fluids, as shown in Figure 2 (*b*) and Figure 3 (*c* and *d*), respectively. The concave is also observed on momentum profiles of the hard-sphere fluids near the wall at higher pore averaged density (Figure 3*e*), since the density near the wall is always bigger than the bulk density. Yet for the real gas, the concave disappears due to the smaller density near the wall comparing to the hard-sphere fluids.

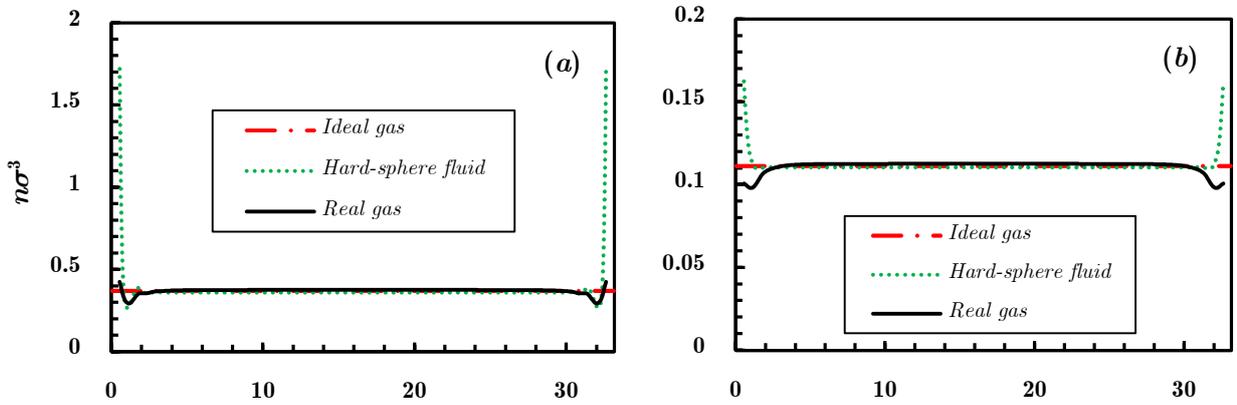

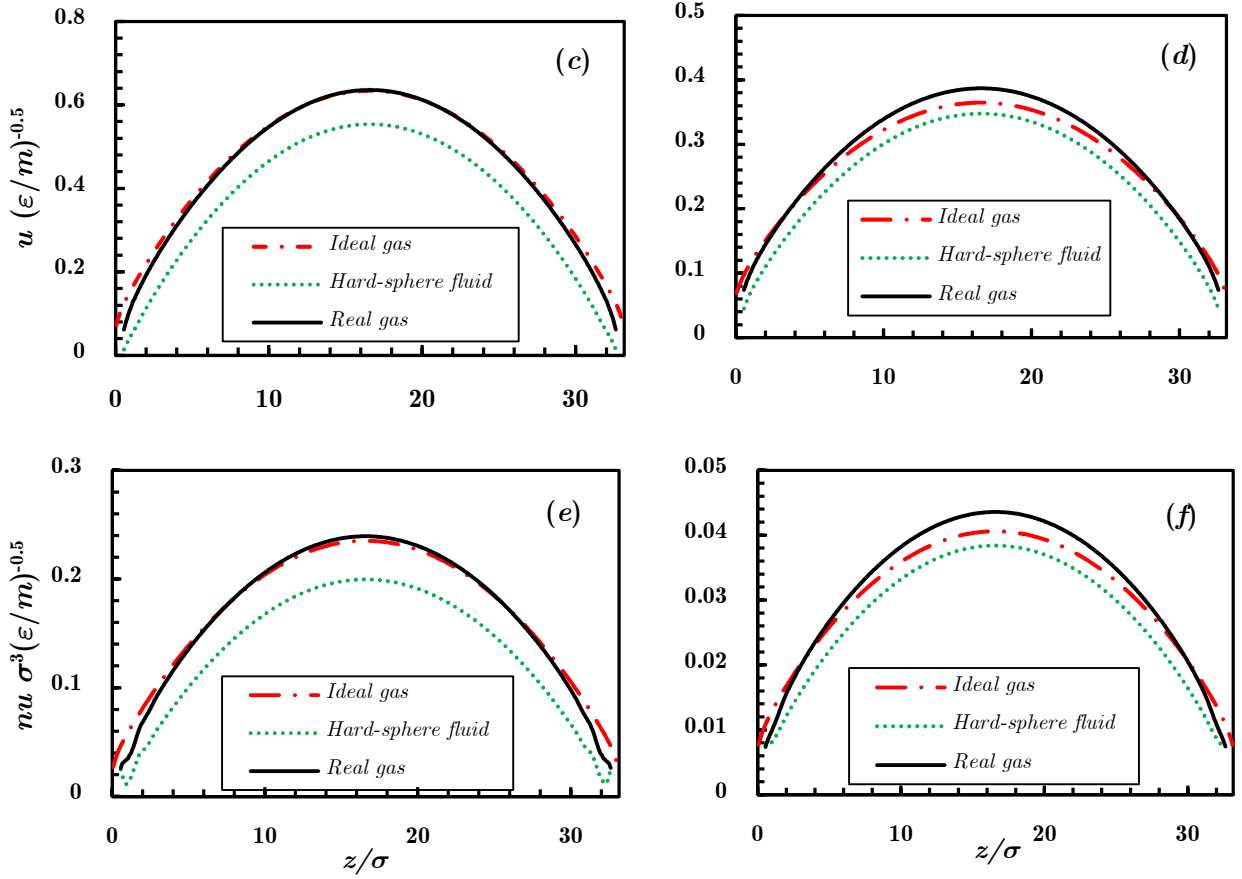

Figure 4: The density (*a* and *b*), velocity (*c* and *d*), and momentum (*e* and *f*) profiles of methane flow in slit channels of $H/\sigma$ =32.16 at the temperature of $T = 2.45 \ \varepsilon/k_B$. The *a*, *c* and *e* are for profiles with the pore averaged density at $n_0 = 0.371 \ \sigma^{-3}$ ($Kn = 0.0183$ for ideal gas, $Kn = 0.0106$ for hard-sphere fluid and real gas), and the *b*, *d* and *f* are for profiles with pore averaged density at $n_0 = 0.111 \ \sigma^{-3}$ ($Kn = 0.0610$ for ideal gas, $Kn = 0.0525$ for hard-sphere fluid and real gas).

The variation of the density, velocity and momentum of fluids confined in larger channels of $H/\sigma = 32.16$ are shown in Figure 4, with other parameters the same as in Figure 3. Similar results are observed for both the high ($n_0 = 0.371 \ \sigma^{-3}$) and low ($n_0 = 0.111 \ \sigma^{-3}$) pore averaged density cases, comparing to gas dynamics in much narrower channels ($H/\sigma = 8.04$ in Figure 3). However, the flow behaviors of real gas is closer to ideal gas ones, while the Enskog theory underestimates the flow velocity and the mass flow rate. This phenomenon can also be observed in Figure 2. With the increase of the channel width, the mass flow rate of the real gas approaches the Boltzmann results faster than the hard-sphere fluids, as the consequence of attractive intermolecular forces pulling the fluids more unitary across the channel. Therefore, the real gas dynamics can be predicted by the Boltzmann equation in macro pores, while the Enskog theory underestimates the flow velocity and the mass flow rate.

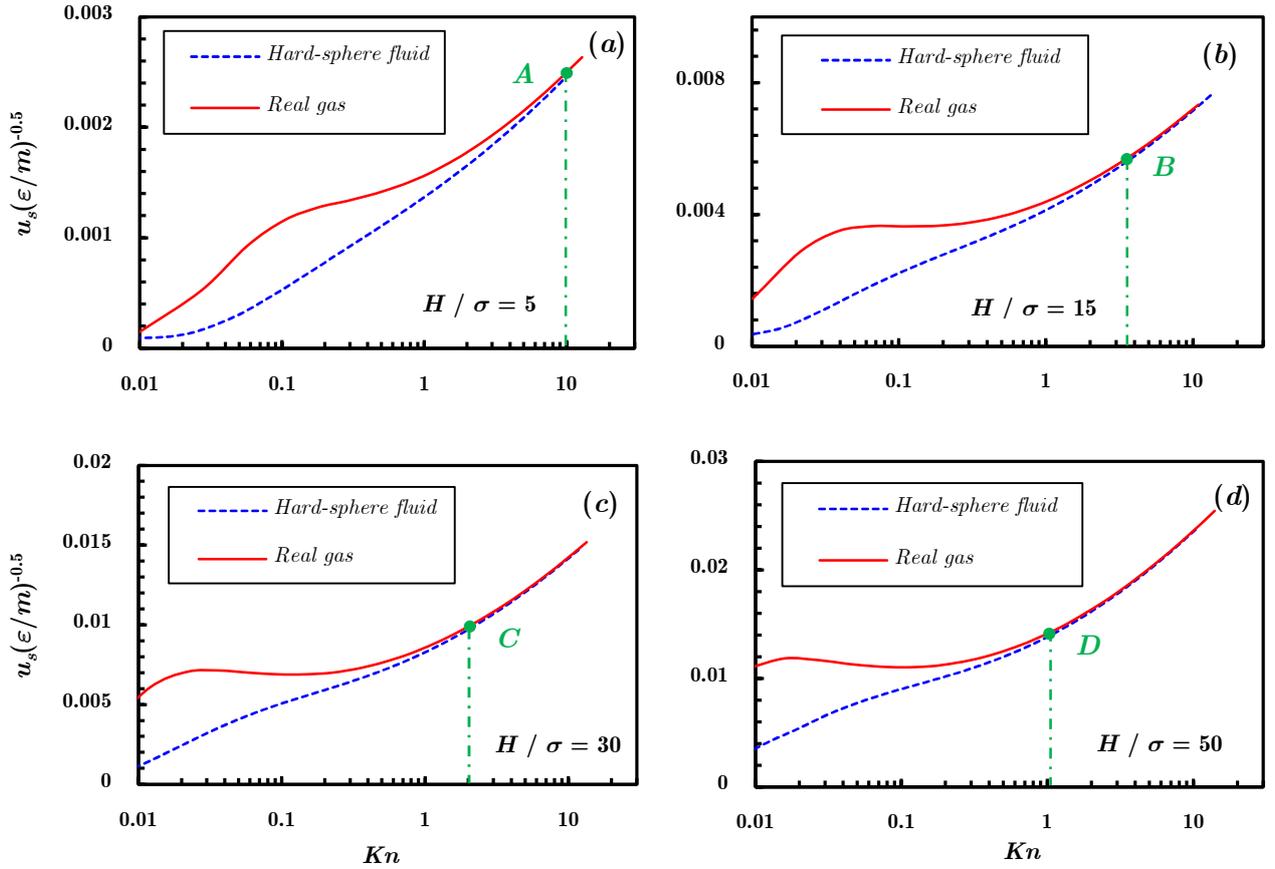

Figure 5: The slip velocity of the hard-sphere fluids and the real gas under different Knudsen number and confinement conditions at the temperature of $T = 2.45\ \varepsilon/k_B$.

In Figure 5, the variation of slip velocity with $Kn$ of hard-sphere fluids and real gas confined in nano-slits of $H/\sigma = 5$, 15, 30 and 50 is displayed. The slip velocity of the hard-sphere fluids increases with the $Kn$, while there is a regime of anomalous slip for the real gas, where the slip velocity barely changes or decreases with the $Kn$ increasing when $H/\sigma \geq 15$, as shown in Figure 5 (*b*), (*c*) and (*d*). The slip velocity of the real gas is larger than that of the hard-sphere fluids when the $Kn$ is below some critical point, namely $A$, $B$, $C$ and $D$ for $H/\sigma = 5$, 15, 30 and 50, respectively. The corresponding $Kn$ of this critical point decreases with the channel width. Therefore, the intermolecular attractive forces need to be taken into account in a broader flow regime when the channel width is smaller. For example, the long-range intermolecular potential affects the slip velocity when $Kn < 1$ for $H/\sigma = 50$, but this critical $Kn$ extends to $Kn = 10$ for $H/\sigma = 5$. For tight gas reservoirs, the flow is mostly located in the slip and transition flow regimes ($Kn < 1$). Consequently, it is necessary to consider the effects of intermolecular attractive forces on gas dynamics when $H/\sigma < 50$.

To understand the gas dynamics in anomalous slip flow regime, the density, velocity, momentum and non-dimensional momentum profiles of fluids confined in slits of $H/\sigma = 30$ are studied for the $Kn = 0.0320$, 0.0827 and 0.1974, respectively. For a fixed channel width, the smaller $Kn$ means the larger solid fraction $\eta$. The density is more inhomogeneous near the wall at smaller $Kn$, and the difference between the hard-sphere fluids and the real gas is also more obvious, as shown in Figure 6 (*a*). The anomalous slip of the real gas and the normal slip of the hard-sphere fluids are observed in Figure 6 (*b*). Although, the slip velocity barely changes with the $Kn$, the central velocity decreases with the increasing $Kn$. Therefore, the mass transfer is more prominent at smaller $Kn$, as we can see from Figure 6 (*b*) and (*c*). Besides, the non-dimensional momentum slip almost coincides near the wall for both the real gas and the hard-sphere fluids (*d*). The mass flow rate (*c* and *d*) of the real gas is larger than that of the hard-sphere fluid for the same $Kn$. This difference decreases with the increasing $Kn$. Although the rarefaction effects enhance the flow by a slip velocity near the wall at large $Kn$, the mass flow rate of dense gas at smaller $Kn$ is much larger due to the larger solid fraction. Therefore, the rarefaction effects should be treated rationally. Otherwise, the role of the rarefaction may be exaggerated.

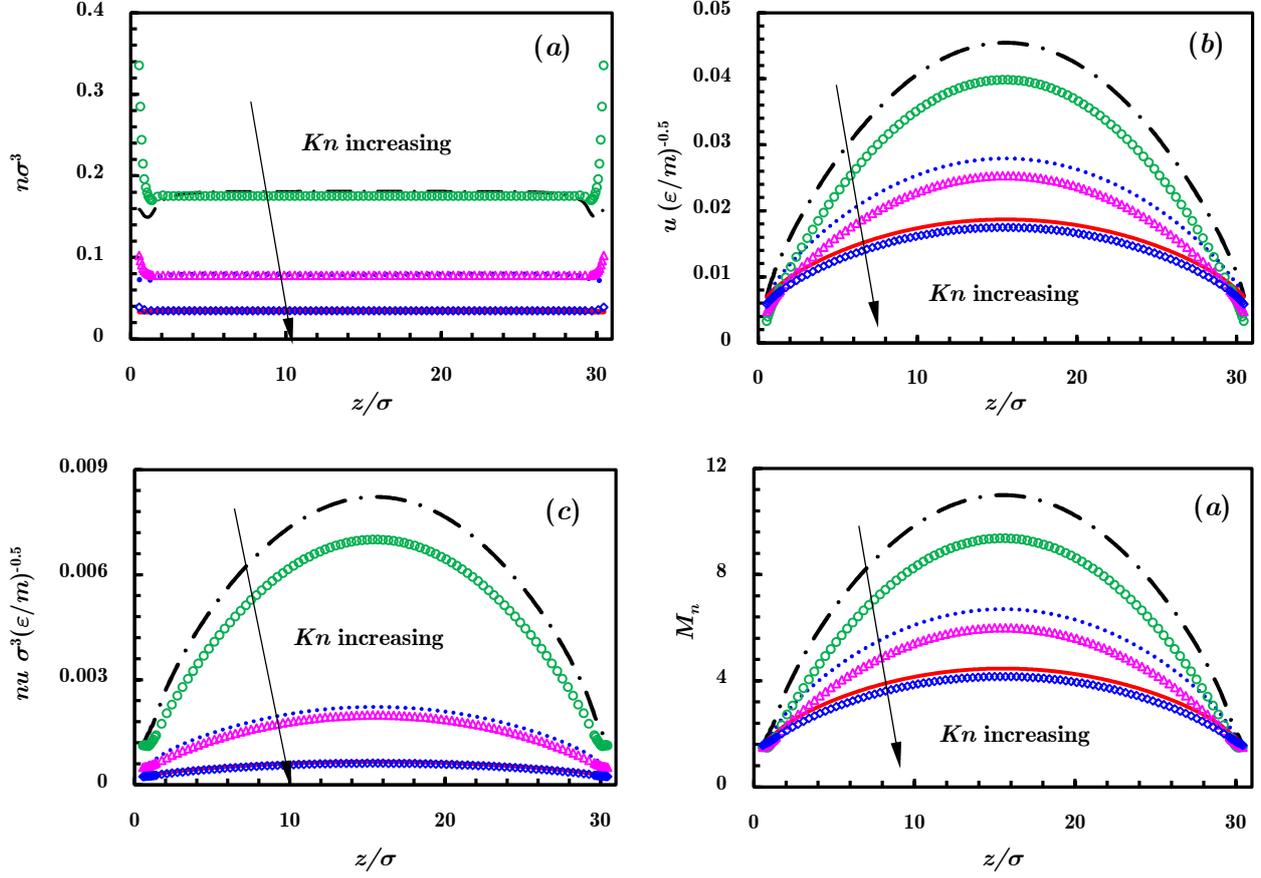

Figure 6: The profiles of (*a*) density, (*b*) velocity, (*c*) momentum and (*d*) non-dimensional momentum for the hard-sphere fluids (symbols) and the real gas (lines) confined in slits of $H/\sigma = 30$ at the temperature of $T = 2.45\ \varepsilon/k_B$. Black dash-dotted lines, $Kn = 0.0320$; blue dash lines, $Kn = 0.0827$; red solid lines, $Kn = 0.1974$; green circles, $Kn = 0.0320$; pink triangles, $Kn = 0.0827$, blue diamonds, $Kn = 0.1974$. The solid fraction are $\eta = 0.0931, 0.0414$ and $0.0184$ for $Kn = 0.0320, 0.0827$ and $0.1974$, respectively.

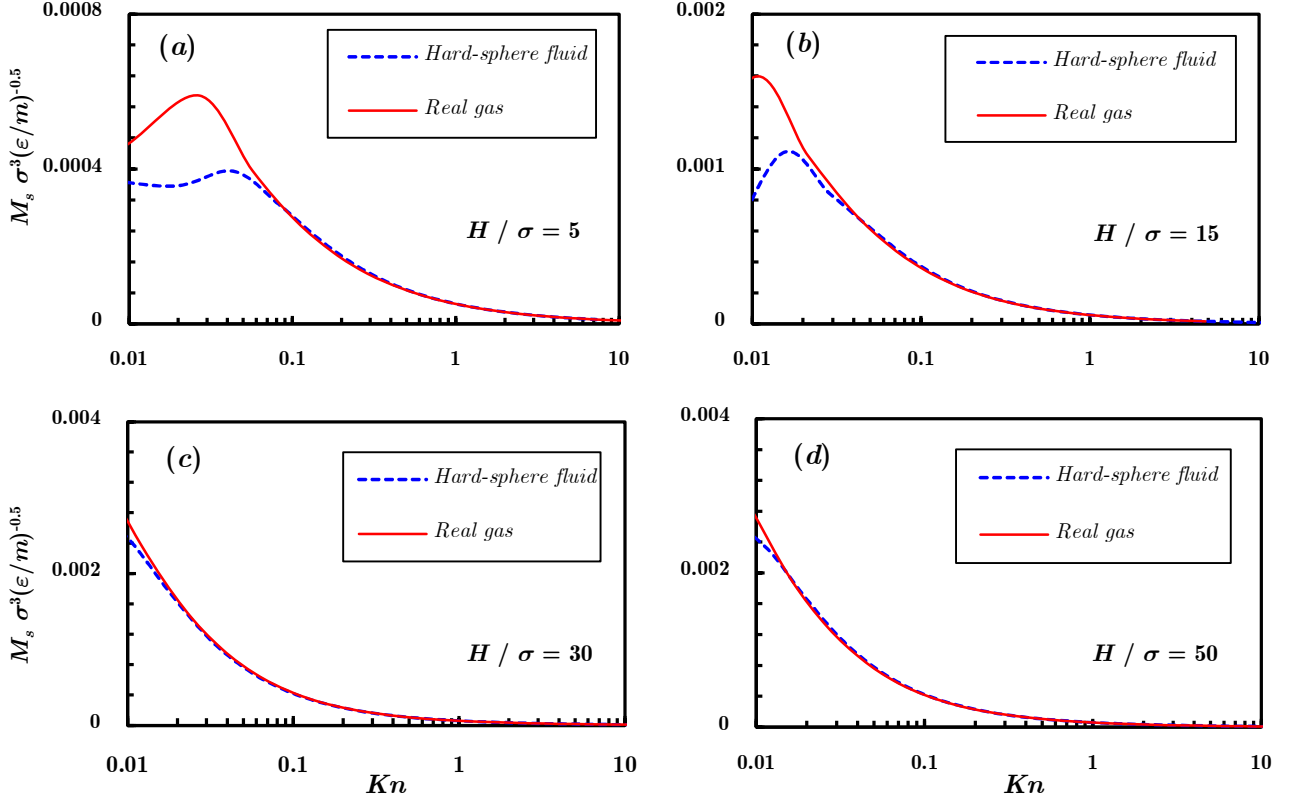

Figure 7: The slip momentum, which is defined as the product of the slip velocity and the local number density, under different Knudsen number and confinement conditions for the hard-sphere fluids and the real gas at the temperature of $T = 2.45\ \varepsilon/k_B$.

Although the slip velocity affects the mass flow rate, it is the momentum that determines the practical amount of mass transfer. The variation of momentum slip with $Kn$ is shown in Figure 7. For fluids under tight confinement ($H/\sigma = 5$), the slip momentum of hard-sphere fluids or the real gas barely changes or increases with the $Kn$ when the $Kn$ is small ($Kn < 0.03$), after which it decreases with the $Kn$ continuously. There is no increase of slip momentum with the increasing $Kn$ in large channels, as we can see from Figure 7 $(c)$ and $(d)$.

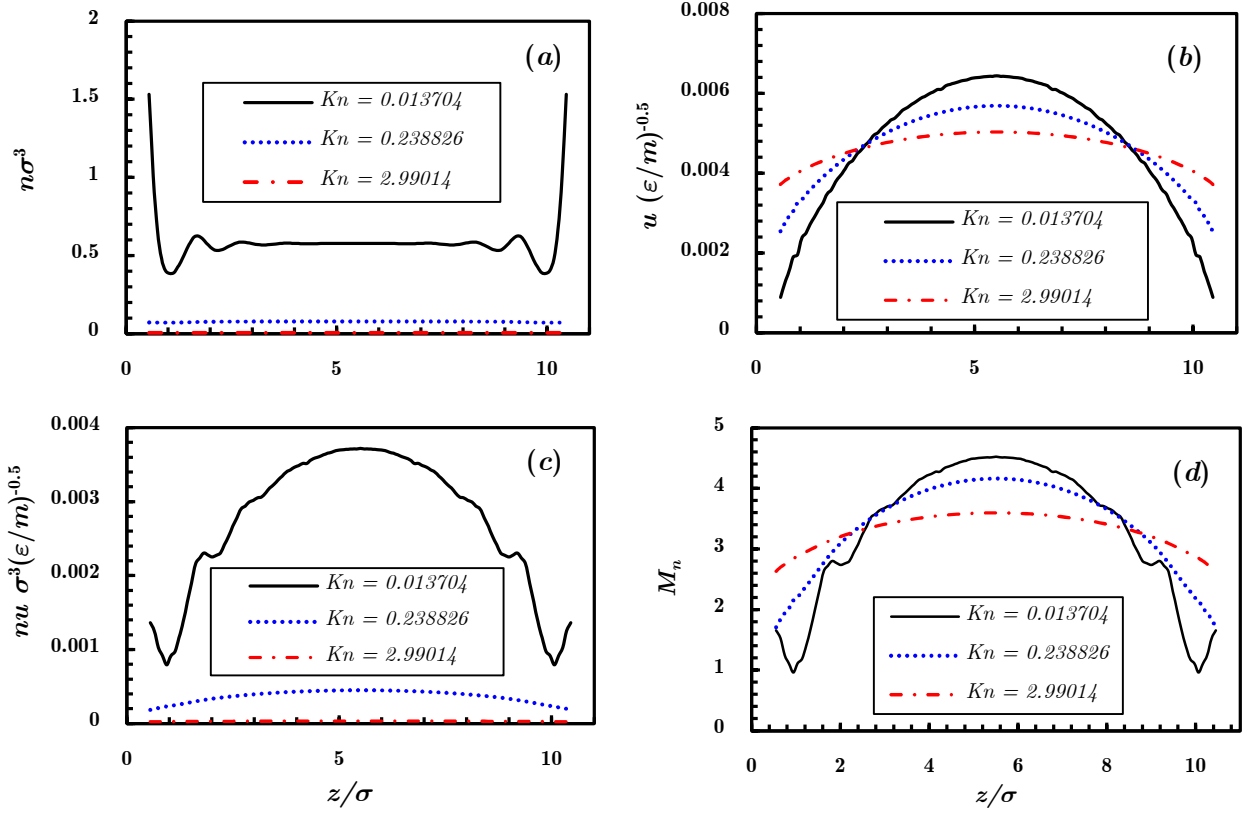

Figure 8: The density (*a*), velocity (*b*), momentum (*c*) and normalized momentum (*d*) profiles of real gas confined in nano-slits of $H/\sigma = 10$ under different Knudsen number conditions at the temperature of $T = 2.45\ \varepsilon/k_B$.

The Knudsen maximum and Knudsen minimum can both be observed in Figure 2 for the hard-sphere fluids and the real gas under confinement ($5 < H/\sigma < 30$). Taking the real gas under the confinement of $H/\sigma = 10$ as an example, the non-dimensional mass flow rates are approximately the same for $Kn = 0.013704, 0.238826, 2.99014$. The density, velocity and momentum profiles across the channel are shown in Figure 8. With the increasing $Kn$, the slip velocity becomes larger and the velocity profile tends to be more plug-like. The rarefaction effects increase the slip velocity, but decrease the maximum velocity. Thus, the overall contribution of the velocity profile at different $Kn$ ($Kn = 0.013704, 0.238826, 2.99014$) on mass flow rate is similar. The trend of the non-dimensional momentum distribution is mainly controlled by the velocity profiles, as we can see from Figure 8 (*d*), although the inhomogeneity of the density at $Kn = 0.013704$ affects the shape of the profile. Yet it is the density that governs the practical amount of mass transfer. The density decreases significantly with the increasing $Kn$ (Figure 8$a$) and the overall contribution of velocity at different $Kn$ to the mass rate is similar (Figure 8$b$), therefore the mass flow rate decreases with the increasing $Kn$, which are $Q = 0.011252$, $0.001522$ and $0.000130$ for $Kn = 0.013704$, $0.238826$ and $2.99014$, respectively. Although the rarefaction effects increase the slip velocity, it is the flow at the small $Kn$ that contributes more mass transfer in porous media.

The variation of slip velocity and momentum profiles with the $Kn$ are displayed in Figure 9. The non-dimensional slip velocity increases with the increasing $Kn$ due to the rarefaction effects (Figure 9$a$), while the slip velocity decreases with the increasing $Kn$ due to the shrinkage of the pore size. Therefore, the pore size is a more important factor that affects the practical amount of mass transfer. This is different from the variation in Figure 5, where the slip velocity and non-dimensional slip velocity have the same trend in changing with the $Kn$. The velocity profiles across the channel under different density conditions ($n_0 = 0.1$ and $n_0 = 0.2$) at approximately the same $Kn$ are shown in Figure 9 ($c$). Although the non-dimensional slip velocity is barely the same for $n_0 = 0.1$ ($H/\sigma = 13.561$) and $0.2$ ($H/\sigma = 5.705$), the maximum non-dimensional velocity and dimensional velocity profile across the channel of $n_0 = 0.1$ are larger than those of $n_0 = 0.2$. Similarly, the non-dimensional mass flow rate decreases with the increasing $Kn$, which further proves our assessment that the flow at smaller $Kn$ contributes to more practical mass flow rate.

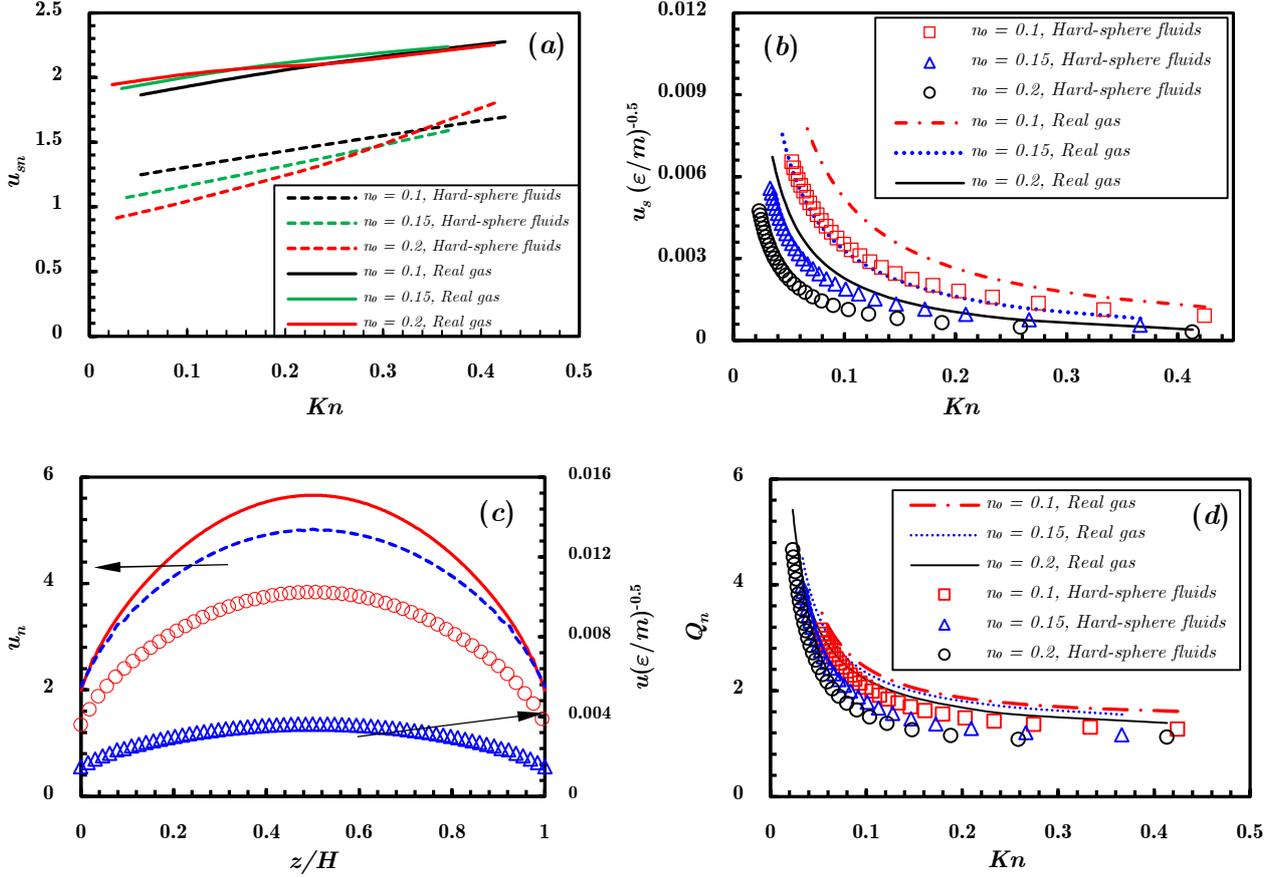

Figure 9: The variation of the normalized slip velocity (a), slip velocity (b) and the normalized flux (d) with the Knudsen number of the hard-sphere fluids and the real gas under different density conditions at the temperature of $T = 2.45 \ \varepsilon/k_B$. The velocity profile of real gas at the $Kn = 0.146$ (c), where the red line represents the non-dimensional velocity at $n_0 = 0.1$ ($H/\sigma = 13.561$), blue dashed line represents the non-dimensional velocity at $n_0 = 0.2$ ($H/\sigma = 5.705$), red circles represent the velocity at $n_0 = 0.1$ ($H/\sigma = 13.561$) and blue triangles represent velocity at $n_0 = 0.2$ ($H/\sigma = 5.705$).

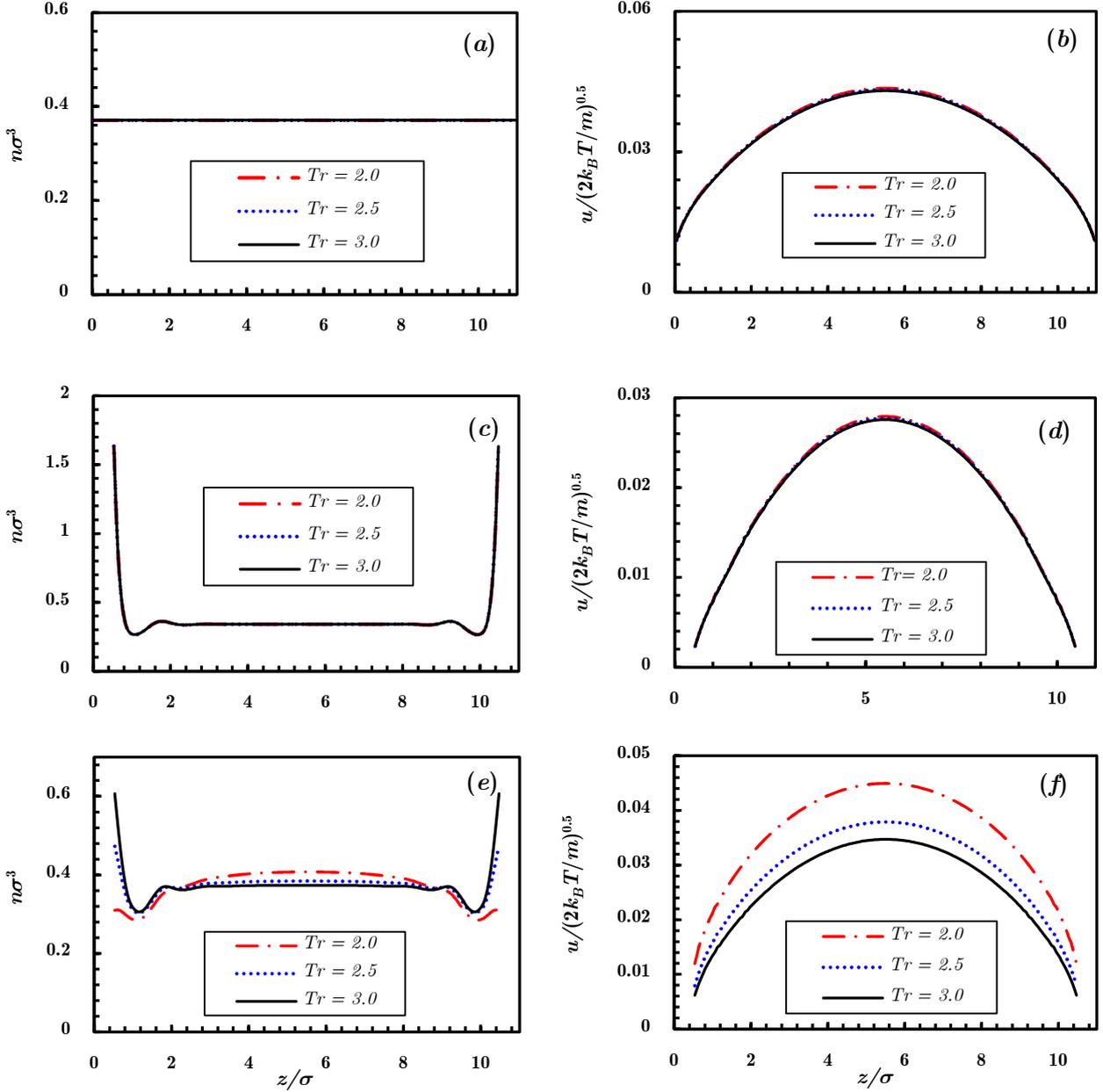

Figure 10: The effects of reduced temperature $T_r = k_B T/\varepsilon$ on the density (*a*, *c* and *e*) and velocity (*b*, *d* and *f*) profiles of ideal gas (*a* and *b*), hard-sphere fluids (*c* and *d*) and real gas (*e* and *f*).

The influence of the temperature on density and velocity profiles of ideal gas, hard-sphere fluids and real gas is investigated in Figure 10. The density and velocity profiles of the ideal gas and hard-sphere fluids are not affected by the temperature (Figure 10 *a*, *b*, *c* and *d*), while the density and velocity profiles of real gas depend significantly on the reduced temperature (Figure 10 *e* and *f*). As the temperature increases, the energy parameter of the fluids $\varepsilon$ decreases, which leads to the decrease of the long-range intermolecular attraction. Thus, the characteristic of the real gas tends to be more like that of the hard-sphere fluids. As shown in Figure 10 $e$, with temperature increasing, the bulk density decreases and more fluid molecules accumulate near the wall due to the volume exclusion. Besides, the velocity decreases with the increasing temperature, and approaches the velocity profile of the hard-sphere fluids comparing Figure 10 ($d$) to Figure 10 ($f$).

## 4 Conclusions

Gas dynamics of the ideal gas, hard-sphere fluids and real gas have been investigated and compared by the Boltzmann-BGK model, the Enskog-BGK model and the simple kinetic model, respectively. The rarefaction effects caused by the non-negligible $Kn$ and the real gas effects caused by the changes of thermodynamic properties are coupled self-consistently in the simple kinetic equation, which is an effective tool to study gas dynamics under confinement at high pressure and temperatures, such as in natural gas reservoirs.

The mass flow rate of the real gas is between that of the ideal gas and hard-sphere fluids when $Kn < 0.3$, while it is almost the same for the three fluid types when $Kn > 1$, where the collisions among molecules are ignorable. Under constant confinement, the anomalous slip is found for the real gas, while the slip velocity of the hard-sphere fluid increases monotonically with the increasing $Kn$. The slip velocity of the real gas is larger than that of the hard-sphere fluids when $Kn$ is small, but increases slower with the $Kn$, and thus, the slip velocity of the hard-sphere fluids and the real gas is the same when $Kn$ is sufficiently large. Although the rarefaction effects enhance the slip velocity, it also decreases the maximum velocity at the centerline. The increasing $Kn$ means a smaller solid fraction or a smaller pore size, both of which lead to smaller mass flow rate. Therefore, it is the flow at the small $Kn$ that dominates the practical mass transfer in porous media, which implies that the role of rarefaction effects in gas transport in natural gas reservoirs should be treated rationally. Meanwhile, the real gas dynamics are significantly affected by the temperature, since the energy among fluid molecules is considered. With the increase of the temperature, the density and velocity profiles of the real gas tend to those of the hard-sphere fluids, as the effect of the long-range attractive potential becomes weaker.

**Acknowledgments**

This work was supported by the National Science Foundation of China (Grant No. 51836003).

[1] Z. Wang, M. Wang, S. Chen, Coupling of high Knudsen number and non-ideal gas effects in microporous media, J. Fluid Mech., 840 (2018) 56-73.

[2] Y. Lan, Z. Yang, P. Wang, Y. Yan, L. Zhang, J. Ran, A review of microscopic seepage mechanism for shale gas extracted by supercritical CO2 flooding, Fuel, 238 (2019) 412-424.

[3] H. Sun, J. Yao, S.-h. Gao, D.-y. Fan, C.-c. Wang, Z.-x. Sun, Numerical study of CO2 enhanced natural gas recovery and sequestration in shale gas reservoirs, International Journal of Greenhouse Gas Control, 19 (2013) 406-419.

[4] K. Wu, Z. Chen, X. Li, J. Xu, J. Li, K. Wang, H. Wang, S. Wang, X. Dong, Flow behavior of gas confined in nanoporous shale at high pressure: Real gas effect, Fuel, 205 (2017) 173-183.

[5] L. Zhang, B. Shan, Y. Zhao, Z. Guo, Review of micro seepage mechanisms in shale gas reservoirs, Int. J. Heat Mass Transfer, 139 (2019) 144-179.

[6] J. Ma, J.P. Sanchez, K. Wu, G.D. Couples, Z. Jiang, A pore network model for simulating non-ideal gas flow in micro- and nano-porous materials, Fuel, 116 (2014) 498-508.

[7] X. Yan, J. Sun, D. Liu, Numerical Simulation of Shale Gas Multiscale Seepage Mechanism-Coupled Stress Sensitivity, Journal of Chemistry, 2019 (2019) 1-13.

[8] Q. Zhang, Y. Su, W. Wang, M. Lu, G. Sheng, Gas transport behaviors in shale nanopores based on multiple mechanisms and macroscale modeling, Int. J. Heat Mass Transfer, 125 (2018) 845-857.

[9] R.-h. Zhang, L.-h. Zhang, R.-h. Wang, Y.-l. Zhao, R. Huang, Simulation of a Multistage Fractured Horizontal Well with Finite Conductivity in Composite Shale Gas Reservoir through Finite-Element Method, Energy & Fuels, 30(11) (2016) 9036-9049.

[10] L. Zhang, B. Shan, Y. Zhao, J. Du, J. Chen, X. Tao, Gas Transport Model in Organic Shale Nanopores Considering Langmuir Slip Conditions and Diffusion: Pore Confinement, Real Gas, and Geomechanical Effects,

Energies, 11(1) (2018).

[11] R. Yang, Z. Huang, W. Yu, G. Li, W. Ren, L. Zuo, X. Tan, K. Sepehrnoori, S. Tian, M. Sheng, A Comprehensive Model for Real Gas Transport in Shale Formations with Complex Non-planar Fracture Networks, Sci Rep, 6 (2016) 36673.

[12] J. Li, A.S. Sultan, Klinkenberg slippage effect in the permeability computations of shale gas by the pore-scale simulations, Journal of Natural Gas Science and Engineering, 48 (2017) 197-202.

[13] M.T. Ho, L. Zhu, L. Wu, P. Wang, Z. Guo, J. Ma, Y. Zhang, Pore-scale simulations of rarefied gas flows in ultra-tight porous media, Fuel, 249 (2019) 341-351.

[14] Y. Wu, P. Tahmasebi, H. Yu, C. Lin, H. Wu, C. Dong, Pore‐Scale 3D Dynamic Modeling and Characterization of Shale Samples: Considering the Effects of Thermal Maturation, Journal of Geophysical Research: Solid Earth, 125(1) (2020).

[15] J. Ren, P. Guo, Z. Guo, Z. Wang, A Lattice Boltzmann Model for Simulating Gas Flow in Kerogen Pores, Transport in Porous Media, 106(2) (2014) 285-301.

[16] J. Wang, L. Chen, Q. Kang, S.S. Rahman, The lattice Boltzmann method for isothermal micro-gaseous flow and its application in shale gas flow: A review, Int. J. Heat Mass Transfer, 95 (2016) 94-108.

[17] J. Zhao, J. Yao, L. Zhang, H. Sui, M. Zhang, Pore-scale simulation of shale gas production considering the adsorption effect, Int. J. Heat Mass Transfer, 103 (2016) 1098-1107.

[18] K. Wu, Z. Chen, X. Li, C. Guo, M. Wei, A model for multiple transport mechanisms through nanopores of shale gas reservoirs with real gas effect–adsorption-mechanic coupling, Int. J. Heat Mass Transfer, 93 (2016) 408-426.

[19] L. Zhang, H. Liang, Y. Zhao, J. Xie, X. Peng, Q. Li, Gas transport characteristics in shale matrix based on multiple mechanisms, Chem. Eng. J., 386 (2020).

[20] H. Tran, A. Sakhaee-Pour, Slippage in shale based on acyclic pore model, Int. J. Heat Mass Transfer, 126 (2018) 761-772.

[21] B. Shan, P. Wang, Y. Zhang, Z. Guo, Discrete unified gas kinetic scheme for all Knudsen number flows. IV. Strongly inhomogeneous fluids, Submitted to Physical Review E,  (2020).

[22] S. Roy, R. Raju, H.F. Chuang, B.A. Cruden, M. Meyyappan, Modeling gas flow through microchannels and nanopores, J. Appl. Phys., 93(8) (2003) 4870-4879.

[23] S. Ansumali, Mean-Field Model Beyond Boltzmann-Enskog Picture for Dense Gases, Communications in Computational Physics, 9(5) (2015) 1106-1116.

[24] S. Chapman, T.G. Cowling, The mathematical theory of non-uniform gases, Cambridge University Press, 1970.

[25] Y. Shi, T.S. Zhao, Z. Guo, Lattice Boltzmann simulation of dense gas flows in microchannels, Phys. Rev. E., 76(1 Pt 2) (2007) 016707.

[26] L. Wu, H. Liu, J.M. Reese, Y. Zhang, Non-equilibrium dynamics of dense gas under tight confinement, J. Fluid Mech., 794 (2016) 252-266.

[27] J.D. Van Der Waals, J.S. Rowlinson, On the continuity of the gaseous and liquid states, Courier Corporation, 2004.

[28] X.Y. He, G.D. Doolen, Thermodynamic foundations of kinetic theory and Lattice Boltzmann models for multiphase flows, J. Stat. Phys., 107(1-2) (2002) 309-328.

[29] M. Sadr, M.H. Gorji, Treatment of long-range interactions arising in the Enskog–Vlasov description of dense fluids, J. Comput. Phys., 378 (2019) 129-142.

[30] A. Frezzotti, L. Gibelli, S. Lorenzani, Mean field kinetic theory description of evaporation of a fluid into vacuum, Phys. Fluids, 17(1) (2005).

[31] M. Kon, K. Kobayashi, M. Watanabe, Method of determining kinetic boundary conditions in net evaporation/condensation, Phys. Fluids, 26(7) (2014).

[32] I. Bitsanis, J.J. Magda, M. Tirrell, H.T. Davis, Molecular dynamics of flow in micropores, J. Chem. Phys., 87(3) (1987) 1733-1750.

[33] Z. Guo, T.S. Zhao, Y. Shi, Simple kinetic model for fluid flows in the nanometer scale, Phys. Rev. E., 71(3 Pt 2A) (2005) 035301.

[34] H. Ted Davis, Kinetic Theory of Flow in Strongly Inhomogeneous Fluids, Chem. Eng. Commun., 58(1-6) (1987) 413-430.

[35] K. Piechór, Discrete velocity models of the Enskog-Vlasov equation, Transport Theory and Statistical Physics, 23(1-3) (1994) 39-74.

[36] J. Karkheck, G. Stell, Kinetic mean‐field theories, J. Chem. Phys., 75(3) (1981) 1475-1487.

[37] P. Tarazona, Free-energy density functional for hard spheres, Phys. Rev. A, 31(4) (1985) 2672-2679.

[38] T.K. Vanderlick, L.E. Scriven, H.T. Davis, Molecular theories of confined fluids, J. Chem. Phys., 90(4) (1989) 2422-2436.

[39] Z. Guo, T.S. Zhao, Y. Shi, Generalized hydrodynamic model for fluid flows: From nanoscale to macroscale, Phys. Fluids, 18(6) (2006).

[40] R.L. Cotterman, B.J. Schwarz, J.M. Prausnitz, Molecular thermodynamics for fluids at low and high densities. Part I: Pure fluids containing small or large molecules, AlChE J., 32(11) (1986) 1787-1798.

[41] X.Y. He, X.W. Shan, G.D. Doolen, Discrete Boltzmann equation model for nonideal gases, Phys. Rev. E., 57(1) (1998) R13-R16.

[42] L.S. Luo, Unified theory of lattice boltzmann models for nonideal gases, Phys. Rev. Lett., 81(8) (1998) 1618-1621.

[43] L.S. Luo, Theory of the lattice boltzmann method: lattice boltzmann models for nonideal gases, Phys. Rev. E., 62(4 Pt A) (2000) 4982-4996.

[44] Z. Guo, R. Wang, K. Xu, Discrete unified gas kinetic scheme for all Knudsen number flows. II. Thermal compressible case, Phys. Rev. E., 91(3) (2015) 033313.

[45] Z. Guo, K. Xu, R. Wang, Discrete unified gas kinetic scheme for all Knudsen number flows: low-speed isothermal case, Phys. Rev. E., 88(3) (2013) 033305.

[46] T. Ohwada, Y. Sone, K. Aoki, Numerical analysis of the Poiseuille and thermal transpiration flows between two parallel plates on the basis of the Boltzmann equation for hard‐sphere molecules, Physics of Fluids A: Fluid Dynamics, 1(12) (1989) 2042-2049.

[47] F. Sharipov, V. Seleznev, Data on Internal Rarefied Gas Flows, J. Phys. Chem. Ref. Data, 27(3) (1998) 657-706.

[48] C. Cercignani, Variational Approach to Boundary-Value Problems in Kinetic Theory, Phys. Fluids, 9(6) (1966).